\documentclass[10pt,twoside,twocolumn,english,aps,manuscript,aps,preprint,show pacs,superscriptaddress]{revtex4}
\usepackage[latin9]{inputenc}
\usepackage{babel}
\usepackage{units}
\usepackage{amstext}
\usepackage{amssymb}
\usepackage{graphicx}
\usepackage[unicode=true,pdfusetitle,
 bookmarks=true,bookmarksnumbered=false,bookmarksopen=false,
 breaklinks=false,pdfborder={0 0 1},backref=false,colorlinks=false]
 {hyperref}

\makeatletter

\DeclareRobustCommand{\greektext}{%
  \fontencoding{LGR}\selectfont\def\encodingdefault{LGR}}
\DeclareRobustCommand{\textgreek}[1]{\leavevmode{\greektext #1}}
\DeclareFontEncoding{LGR}{}{}
\DeclareTextSymbol{\~}{LGR}{126}
\newcommand{\lyxmathsym}[1]{\ifmmode\begingroup\def\b@ld{bold}
  \text{\ifx\math@version\b@ld\bfseries\fi#1}\endgroup\else#1\fi}

\@ifundefined{textcolor}{}
{%
 \definecolor{BLACK}{gray}{0}
 \definecolor{WHITE}{gray}{1}
 \definecolor{RED}{rgb}{1,0,0}
 \definecolor{GREEN}{rgb}{0,1,0}
 \definecolor{BLUE}{rgb}{0,0,1}
 \definecolor{CYAN}{cmyk}{1,0,0,0}
 \definecolor{MAGENTA}{cmyk}{0,1,0,0}
 \definecolor{YELLOW}{cmyk}{0,0,1,0}
}

\makeatother

\begin{document}

\title{$^{75}$As NMR local probe study of magnetism in (Eu$_{1-x}$K$_{x}$)Fe$_{2}$As$_{2}$}

\author{Tusharkanti Dey}

\email[Email: ]{tusdey@gmail.com}

\altaffiliation[Present address: ]{IFW Dresden, 01171 Dresden, Germany}

\affiliation{Department of Physics, Indian Institute of Technology Bombay, Powai,
Mumbai 400076, India}

\author{P. Khuntia}

\altaffiliation[Present address: ]{Max-Planck Institute for Chemical Physics of Solids, D-01187 Dresden, Germany}

\affiliation{Department of Physics, Indian Institute of Technology Bombay, Powai,
Mumbai 400076, India}

\author{A.V. Mahajan}

\affiliation{Department of Physics, Indian Institute of Technology Bombay, Powai,
Mumbai 400076, India}

\author{Anupam}

\affiliation{Department of Physics, Indian Institute of Technology Kanpur, Kanpur
208016, India}

\author{Z. Hossain}

\affiliation{Department of Physics, Indian Institute of Technology Kanpur, Kanpur
208016, India}
\begin{abstract}
$^{75}$As NMR measurements were performed as a function of temperature
and doping in Eu$_{1-x}$K$_{x}$Fe$_{2}$As$_{2}$ ($x=0.5,0.7$)
samples. The large Eu$^{2+}$ moments and their fluctuations are found
to dominate the $^{75}$As NMR properties. The $^{75}$As nuclei close
to the Eu$^{2+}$ moments likely have a very short spin-spin relaxation
time ($T_{2}$) and are wiped out of our measurement window. The $^{75}$As
nuclei relatively far from Eu$^{2+}$ moments are probed in this study.
Increasing the Eu content progressively decreases the signal intensity
with no signal found for the full-Eu sample ($x=0$). The large $^{75}$As
NMR linewidth arises from an inhomogeneous magnetic environment around
them. The spin lattice relaxation rate ($1/T_{1}$) for $x=0.5$ and
$0.7$ samples is nearly independent of temperature above $100$\,K
and results from a dipolar coupling to paramagnetic fluctuations of
the Eu$^{2+}$ moments. The behavior of $1/T_{1}$ at lower temperatures
has contributions from the antiferromagnetic fluctuations of the Eu$^{2+}$
moments as also the fluctuations intrinsic to the FeAs planes and
from superconductivity.
\end{abstract}


\maketitle

\section{introduction}

The interest in high temperature superconductivity experienced a resurgence
after the discovery of superconductivity in LaFeAs(O$_{1-x}$F$_{x}$)
with a superconducting transition temperature $T_{\mathrm{c}}=26$\,K
\cite{Kamihara-JACS-130-2008}. So far, hundreds of materials in the
iron pnictide family have been found to be superconducting \cite{Johnston-ReviewFeAs-2010,Stewart-RevModPhy-2011},
some with transition temperatures ranging upto $56$\,K \cite{Wu-JPCM-21-2009}.
These materials broadly belong to $4$ groups based on their crystal
structure. The quaternary `$1111$' compounds with chemical formula
RFeAsO where R is a rare earth element \cite{Kamihara-JACS-130-2008,Chen-Nature-453-2008},
ternary arsenides (`$122$') AFe$_{2}$As$_{2}$ with A = Ba, Sr,
Ca and Eu \cite{Rotter-PRL-101-2008,Jeevan-PRB-78-2008 (EuK),Krellner-PRB-2008(SrFe2As2),Ni-PRB-2008(CaFe2As2)},
`$111$' compounds (Li/Na)FeAs \cite{Tapp-PRB-78-2008,Parker-ChemComm-2009-NaFeAs}
and the `$11$' binary chalcogenides FeSe$_{1-x}$ \cite{Hsu-PNAS-105-2008}.
Among them, `$122$' series materials are studied most because of
their rich phase diagram although the highest $T_{C}$ obtained is
relatively lower than for `$1111$' series materials.

Like other parent compounds in the `$122$' series iron arsenide family,
EuFe$_{2}$As$_{2}$ crystallizes in ThCr$_{2}$Si$_{2}$-type structure
($I4/mmm$) and undergoes a spin density wave (SDW) transition related
to the Fe moments at $\sim200$\,K, confirmed by heat capacity and
resistivity measurements \cite{Jeevan-PRB-78-2008,Ren-PRB-78-2008}.
But EuFe$_{2}$As$_{2}$ is a very special member in the `$122$'
family as it contains two different magnetic atoms. It is the only
known member of the `$122$' family which contains $4f$ electrons.
The Eu$^{2+}$ ions have a large magnetic moment of $7\mu_{\mathrm{B}}$
and they order ferromagnetically within the \textit{$ab$} plane and
antiferromagnetically with neighboring planes along the $c$-axis
at $20$\,K \cite{Jeevan-PRB-78-2008,Ren-PRB-78-2008}. Besides multiple
magnetic transitions, appearance of superconductivity is reported
in EuFe$_{2}$As$_{2}$ by hole/electron/isovalent doping \cite{Jeevan-PRB-78-2008 (EuK),Jiang-PRB-80-2009,Ren-PRL-102-2009},
and also by applying external pressure \cite{Miclea-PRB-79-2009}.
It is also reported that magnetic ordering of Eu$^{2+}$ moments and
superconductivity coexist in this system at low temperature \cite{Anupam-JPCM-21-2009,Ren-PRL-102-2009}.
This material is a good candidate to study the interplay between superconductivity
and Eu$^{2+}$ moments ordering. 

Jeevan \textit{et al.} \cite{Jeevan-PRB-78-2008 (EuK)} first reported
that $50\%$ K substitution in EuFe$_{2}$As$_{2}$ system suppresses
the SDW transition and in turn gives rise to high-temperature superconductivity
below $32$\,K. They also found a signature of Eu$^{2+}$ magnetic
ordering from their specific heat data. From $^{57}$Fe and $^{151}$Eu
Mossbauer spectroscopic study on Eu$_{0.5}$K$_{0.5}$Fe$_{2}$As$_{2}$,
Anupam \textit{et al.} \cite{Anupam-JPCM-21-2009} found no signature
of a SDW transition but Eu$^{2+}$ ordering is found at $10$\,K.
Subsequently, they constructed the phase diagram of the Eu$_{1-x}$K$_{x}$Fe$_{2}$As$_{2}$
($0\leq x\leq1$) system \cite{Anupam-JPCM-23-2011}. The SDW transition
corresponding to the Fe-sublattice coexists with superconductivity
with a lower $T_{C}=5.5$\,K in the underdoped sample ($x=0.15$).
As the doping percentage increases ($x=0.3$) the SDW transition vanishes
and superconductivity appears at relatively higher $T_{C}=20$\,K.
The maximum $T_{C}=32$\,K is obtained for $x=0.5$. Coexistence
of Eu$^{2+}$ magnetic ordering with superconductivity is sustained
upto $x=0.6$. 

Iron arsenide superconductors have been extensively studied using
NMR to obtain a deeper understanding of various physics issues \cite{Ishida-review,Ning-PRL-104-2010,Dey-JPCM-23-2011,Grafe-PRL-101-2008,Nakai-PRB-81-2010}.
In this Eu$_{1-x}$K$_{x}$Fe$_{2}$As$_{2}$ system also $^{75}$As
NMR measurements could be a useful probe to understand the interplay
of magnetic ordering and superconductivity. Till date only two publications
on NMR in EuFe$_{2}$As$_{2}$ and related systems have been reported
\cite{Sarkar-JPCM-24-2012,Guguchia-PRB-83-2011}. In the former, $^{75}$As
NMR measurements are reported on single-crystal of EuFe$_{1.9}$Co$_{0.1}$As$_{2}$
which does not show superconductivity but undergoes an SDW transition
at $120$\,K \cite{Guguchia-PRB-83-2011}. The latter reports $^{75}$As
NMR measurements on a Eu$_{0.2}$Sr$_{0.8}$(Fe$_{0.86}$Co$_{0.14}$)$_{2}$As$_{2}$
single crystal, which shows a superconducting transition below $20$\,K
\cite{Sarkar-JPCM-24-2012}. However no NMR study has been reported
on the Eu$_{1-x}$K$_{x}$Fe$_{2}$As$_{2}$ series which could be
useful to understand how the presence of Eu$^{2+}$ moments affects
the magnetic properties of the material. We have performed a $^{75}$As
NMR study on different doping concentrations ($x=0,0.38,0.5,0.7$)
in (Eu$_{1-x}$K$_{x}$)Fe$_{2}$As$_{2}$ to understand the evolution
of NMR parameters with doping and temperature. Our bulk measurements
show appearance of superconductivity for the $x=0.38,0.5$ and $0.7$
samples with highest $T_{C}=32$\,K for the $x=0.5$ sample. For
the $x=0$ sample, no superconductivity is found but the SDW transition
is present at $197$\,K. However, the SDW transition is suppressed
for the other three samples ($x=0.38,0.5,0.7$). The ordering of the
Eu$^{2+}$ moments is observed at $20$\,K, $10$\,K and $7$\,K
for the $x=0,0.38$ and $0.5$ samples, respectively but no ordering
of the Eu$^{2+}$ moments is found for $x=0.7$. The magnetic properties
of the samples are dominated by the large moment of Eu$^{2+}$ ions.
The $^{75}$As nuclei close to the Eu$^{2+}$ moments have a very
short spin-spin relaxation time ($T_{2}$) driving them out of the
measurement window. This results in a drop in the integrated signal
intensity with increasing Eu-content and no signal was observed with
our measurement conditions for the full Eu ($x=0$) sample. Presence
of unequal environments for the $^{75}$As nuclei (some near Eu moments
and some not) gives rise to broad and asymmetric spectra. The temperature
dependence of the $^{75}$As NMR shift (as determined from the centre-of-gravity)
has a Curie-Weiss behavior reflecting the paramagnetism of the Eu$^{2+}$
spins. From our measurements ($x=0.5,0.7$), we find that the $1/T_{1}$
is nearly temperature independent above $100$\,K (and much enhanced
compared to analogous compositions not containing Eu). This is due
to the paramagnetic fluctuations of the Eu$^{2+}$ moments. The $T$-dependence
at lower temperatures may be related to the intrinsic behavior of
the FeAs planes.

\section{experimental details}

Polycrystalline samples of Eu$_{1-x}$K$_{x}$Fe$_{2}$As$_{2}$ ($x=0,0.38,0.5,0.7$)
were prepared by solid state reaction methods as detailed in Ref.
\cite{Anupam-JPCM-23-2011}. 

We have performed $^{75}$As NMR measurements on (Eu$_{1-x}$K$_{x}$)Fe$_{2}$As$_{2}$
($x=0,0.38,0.5,0.7$) samples using a Tecmag pulse spectrometer in
a magnetic field of $93.954$\,kOe obtained inside a room-temperature
bore Varian superconducting magnet. For variable temperature measurements,
we have used an Oxford continuous flow cryostat. Liquid nitrogen and
liquid helium were used as coolants in the temperature range $80-300$\,K
and $4-80$\,K, respectively. For NMR measurements, we tried to align
the powder samples by mixing with Stycast $1266$ epoxy and then curing
overnight in an external magnetic field $H=93.954$\,kOe. Our measurements
indicate that although the spectral width is reduced significantly,
the samples are not oriented fully. The $^{75}$As has a nuclear spin
$I=\nicefrac{3}{2}$ ($100\%$ natural abundance) with gyromagnetic
ratio $\nicefrac{\gamma}{2\pi}=7.2919$\,MHz/T. Frequency sweep spectra
at a few temperatures were constructed by plotting the integral of
the spin echo (resulting from a $\nicefrac{\lyxmathsym{\textgreek{p}}}{2}-\tau-\nicefrac{\lyxmathsym{\textgreek{p}}}{2}$
pulse sequence with $\tau\geq35\mu\mathrm{s}$) at different transmitter
frequencies. Spin-lattice relaxation was measured using the saturation
recovery method.

\section{results and discussions}

Phase purity of the samples was confirmed by powder x-ray diffraction
(XRD) measurements. Detailed analysis of the XRD data (reported in
Ref. \cite{Anupam-JPCM-23-2011}) shows that lattice parameter \textit{a}
decreases while \textit{c} and the unit cell volume increase with
increasing doping percentage $x$.

\begin{figure}
\centering{}\includegraphics[scale=0.32]{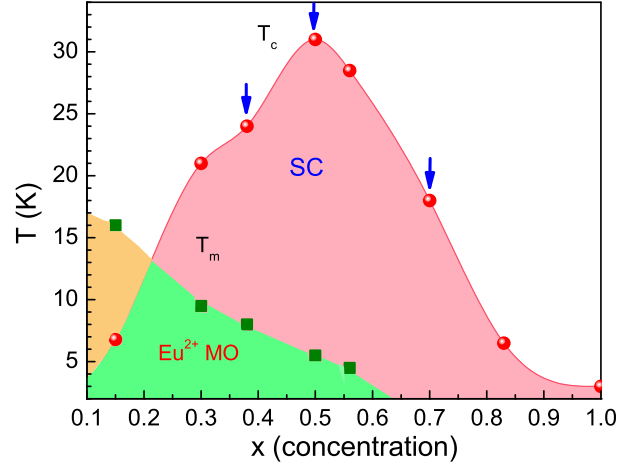}\caption{\label{fig:Phasediagram} The phase diagram of (Eu$_{1-x}$K$_{x}$)Fe$_{2}$As$_{2}$
samples showing the superconducting transition temperature ($T_{C}$)
and Eu$^{2+}$ magnetic ordering temperature ($T_{m}$) as a function
of K content $x$. The figure is adapted from Ref. \cite{Anupam-JPCM-23-2011}.
The compositions used in the present work are indicated with arrows.}
\end{figure}

\begin{figure}
\begin{centering}
\includegraphics[scale=0.32]{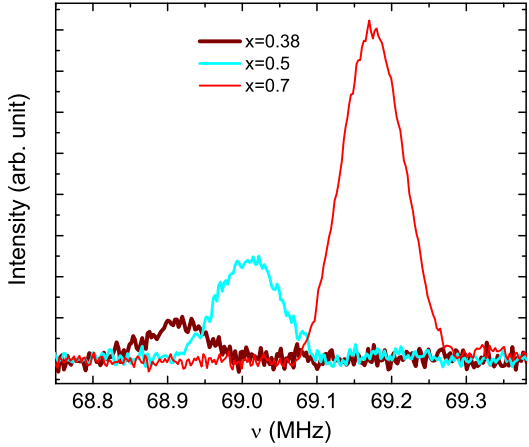}
\par\end{centering}

\caption{\label{fig:SpectraComparison} Fourier transform of the spin echo
for different compositions at room temperature. The transmitter frequency
was adjusted to the position of the peak in each case. Other than
that, these measurements are done with identical conditions (coil,
filling factor, etc.) and the intensities are normalized to the number
of $^{75}$As nuclei present (depends on the molecular weight of a
composition and the mass taken) in the sample. }
\end{figure}

Magnetic susceptibility and resistivity of the samples were measured
as a function of temperature and are reported in Ref. \cite{Anupam-JPCM-23-2011}.
A phase diagram showing the Eu$^{2+}$ moments ordering and the superconducting
transition temperature as a function of doping percentage is shown
in Fig. \ref{fig:Phasediagram}. This phase diagram is adapted from
Ref. \cite{Anupam-JPCM-23-2011} based on measurements on the same
samples as used in the current study. Four of these samples ($x=0,0.38,0.5,0.7$)
are studied in this present work. 

\begin{figure}
\begin{centering}
\includegraphics[scale=0.46]{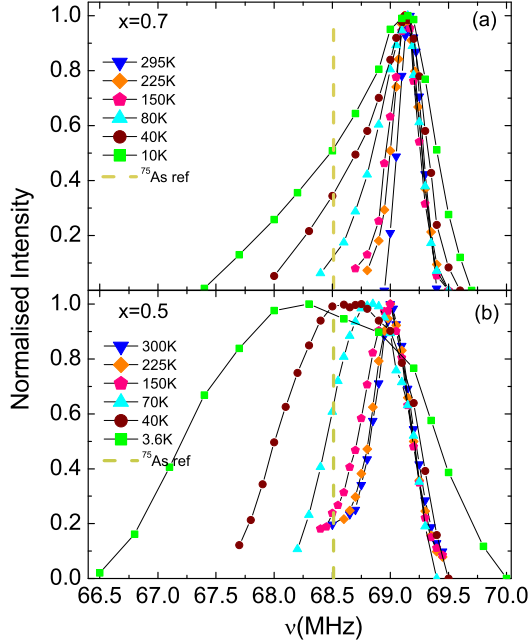}
\par\end{centering}

\caption{\label{fig:CombinedSpectra} $^{75}$As NMR spectra of $x=0.5$ and
$0.7$ samples measured at different temperatures obtained from an
integral of the echo at different frequencies. The $^{75}$As reference
line is shown as dashed line.}
\end{figure}

Following these basic characterizations, we investigated the normal
state properties using $^{75}$As NMR as a local probe. Since the
$^{75}$As nucleus ($I=3/2$) is not at a site of cubic symmetry,
a non-zero electric field gradient (EFG) will be present at the $^{75}$As
site. The interplay between this EFG and the quadrupole moment of
$^{75}$As nucleus will create one central line ($-1/2\rightarrow1/2$
transition) along with satellite peaks ($-3/2\rightarrow-1/2$ and
$3/2\rightarrow1/2$ transitions) on either side of the central line
in the spectra. A powder pattern is expected for randomly oriented
polycrystalline samples. In the present case, we could obtain a partial
alignment with a siginificant reduction in the width of the central
line. We report our measurements on the central line of these partially
aligned samples. We have noticed that as the doping percentage $x$
decreases, the signal intensity gets reduced and no signal is found
for the $x=0$ sample. To quantify this issue, we have measured the
spin-echo for different compositions at their respective spectral
peak positions under identical measurement parameters. The Fourier
transforms of the spin-echos after normalizing to the number of $^{75}$As
nuclei present in the sample are shown in Fig. \ref{fig:SpectraComparison}.
It is clearly seen that the signal intensity sharply falls with decreasing
$x$ i.e increasing Eu content. The ratio of the integrated intensity
at room temperature (spectral width is taken into account) for $x=0.7$
and $0.5$ samples is roughly $1:0.5$. Figure \ref{fig:SpectraComparison}
suggests that the integrated intensity for the $x=0.38$ sample would
be much lower than the other two samples.

\begin{figure}
\centering{}\includegraphics[scale=0.3]{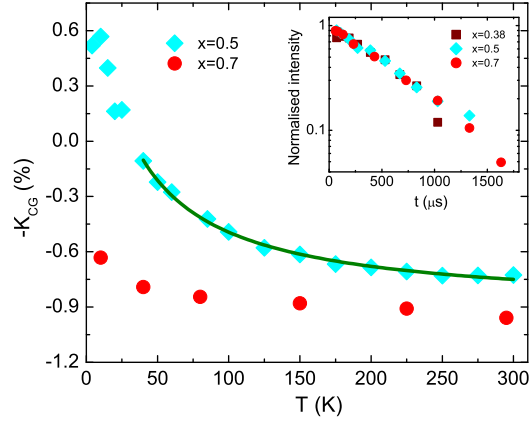}\caption{\label{fig:Shift}$^{75}$As NMR shift (determined using the cg) is
shown as a function of temperature for the $x=0.5$ and $0.7$ samples.
The green solid line is a fit to CW law. Inset: Spin-spin relaxation
($T_{2}$) decay curves for $x=0.38,0.5$ and $0.7$ samples measured
at room temperature.}
\end{figure}

In (Eu$_{1-x}$K$_{x}$)Fe$_{2}$As$_{2}$ samples, the $2a$ site
is occupied by Eu$^{2+}$ and K$^{+}$ ions. In general, the two types
of ions occupy the sites statistically. While the Eu$^{2+}$ has a
large moment, K$^{+}$ does not carry a moment. There will be a fraction
of $^{75}$As nuclei which are close to the Eu$^{2+}$ moments while
some others will be far form Eu$^{2+}$ moments along with all intermediate
possibilities. The $^{75}$As nuclei situated in close proximity to
the Eu$^{2+}$ moments are likely to have a very short spin-spin relaxation
time $T_{2}$ as also a short spin-lattice relaxation time $T_{1}$.
Guguchia \textit{et al.} \cite{Guguchia-PRB-83-2011} measured $^{75}$As
$T_{2}$ for EuFe$_{1.9}$Co$_{0.1}$As$_{2}$ as a function of temperature.
The $T_{2}$ at room temperature was found to be $\sim32\mu\mathrm{s}$
and further decreases with decreasing temperature giving rise to a
wipeout effect with nearly $90\%$ of the signal lost by $100$\,K.
A similar situation is expected for our $x=0$ sample since the composition
is nearly the same. Further, in our measurements the time delay between
the echo forming pulses is larger than $35\mu\mathrm{s}$ which leads
to a significant wipeout even at room temperature. Even in our other
compositions, the $^{75}$As nuclei close to Eu$^{2+}$ moments are
expected to have a short $T_{2}$ and are wiped out of our measurement
window. Consequently, our measurements pertain to nuclei which are
relatively far from Eu$^{2+}$ moments. This fraction is seen to grow
with decreasing europium content. That we are measuring only nuclei
which are far from Eu$^{2+}$ moments is evident from our $T_{2}$
measurements (shown in the inset of Fig. \ref{fig:Shift}). For all
the three samples ($x=0.38,0.5,0.7$) $T_{2}$ at room temperature
is $\sim600\mu\mathrm{s}$. Because of a very low signal intensity
for the $x=0.38$ sample, we have performed temperature dependent
study on $x=0.5$ and $0.7$ samples only.

A wide distribution of magnetic environments at the $^{75}$As sites
gives rise to very broad spectra as shown in Fig. \ref{fig:CombinedSpectra}.
This distribution is greater for the $x=0.5$ composition than the
$x=0.7$ composition, as might be expected. The central line of the
spectra at different temperatures are constructed by plotting the
spin-echo intensity as a function of measuring frequency. Evolution
of the spectra with temperature are shown in Fig. \ref{fig:CombinedSpectra}(b)
and Fig. \ref{fig:CombinedSpectra}(a) for $x=0.5$ and $x=0.7$ samples,
respectively.

Spectra at room temperature for all the three samples are shifted
positively from the reference frequency. The shift at room temperature
is dependent on the Eu-content and varies linearly with $x.$ Spectra
for the $x=0.5$ sample (Fig. \ref{fig:CombinedSpectra}(b)) are almost
symmetric at all temperatures in contrast to that for the $x=0.7$
sample. A possible explanation is that in this case, there is an equal
distribution of $^{75}$As nuclei with more or less Eu neighbours
(assuming statistical occupancy of Eu and K). Further we have studied
the variation of $^{75}$As NMR spectra and shift with temperature.
We have considered the centre of gravity (cg) of the spectra to calculate
the shift ($K_{\mathrm{cg}}$). The temperature dependence of shift
is plotted in Fig. \ref{fig:Shift}. Although the shift is positive
at room temperature, it decreases with decreasing temperature and
becomes negative at low temperature. This temperature variation of
$K_{\mathrm{cg}}$ follows the CW law {[}$K_{cg}(T)=K_{0}+C/(T-\theta)${]}
in the temperature range $40-300$\,K and yields $K_{0}=0.91\%$
and $\theta=-24$\,K (see Fig. \ref{fig:Shift}). The $T$-independent
part $K_{0}$ could be partly arising from the orbital shift while
a part of it can also be due to the intrinsic susceptibility of the
FeAs planes. 

On the other hand, for the $x=0.7$ sample has a weaker temperature
dependence of $K_{cg}$ (Fig. \ref{fig:Shift}). The spectrum is symmetric
at room temperature but becomes asymmetric at lower temperatures (Fig.
\ref{fig:CombinedSpectra}(a)). In this sample, since K percentage
is much more than Eu, there will be a large fraction of $^{75}$As
nuclei far from Eu$^{2+}$ moments. Their resonance frequency does
not shift much with temperature while the ones near Eu moments shift
in a CW manner with decreasing temperature which explains the increasing
asymmetry at lower temperatures. Finally, the magnetic properties
of the samples are dominated by the Eu$^{2+}$ moments and the $^{75}$As
nuclei have a negative hyperfine coupling with the Eu$^{2+}$ moments,
similar as found in Ref. \cite{Guguchia-PRB-83-2011}.

\begin{figure}
\centering{}\includegraphics[scale=0.32]{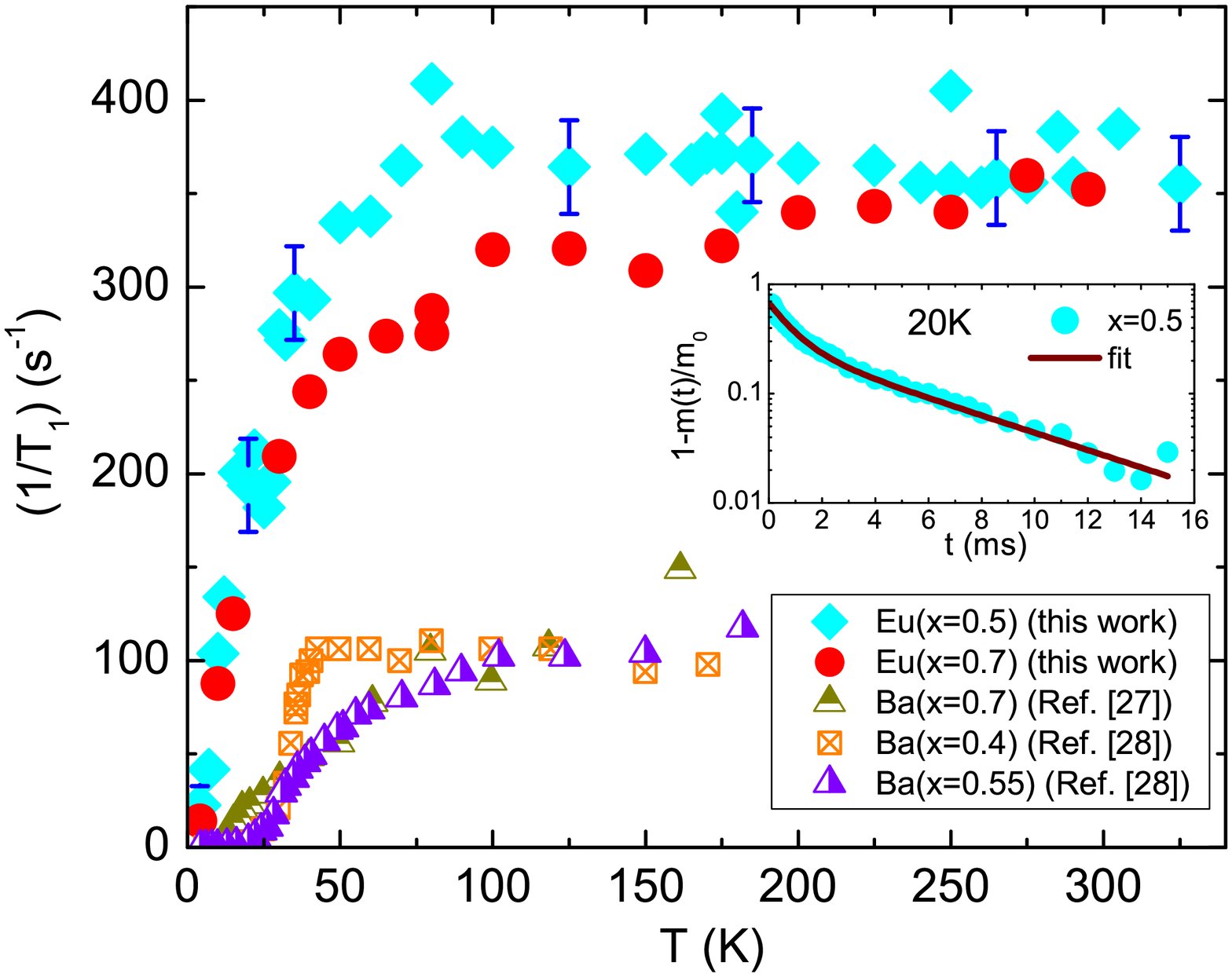}\caption{\label{fig:T1inv} Spin-lattice relaxation rate ($1/T_{1}$) for $x=0.5$
and $x=0.7$ samples are shown as a function of temperature. Relaxation
rate for Ba$_{0.3}$K$_{0.7}$Fe$_{2}$As$_{2}$ {[}Ba($x=0.7$){]}
taken from Ref. \cite{Zhang-PRB-81-2010-Ba0.3K0.7Fe2As2-NMR} and
for Ba$_{0.6}$K$_{0.4}$Fe$_{2}$As$_{2}$ {[}Ba($x=0.4$){]} and
Ba$_{0.45}$K$_{0.55}$Fe$_{2}$As$_{2}$ {[}Ba($x=0.55$){]} taken
from Ref. \cite{Fukazawa-PhysicaC-470-2010} are also shown for comparison.
Inset: A representative spin-lattice relaxation recovery curve for
the $x=0.5$ sample measured at $20$\,K is shown.}
\end{figure}

To study the low energy spin dynamics, we have measured the spin-lattice
relaxation rate ($1/T_{1}$) as a function of temperature by the saturation
recovery method for both the samples ($x=0.5,0.7$). Since the spectra
are very broad, it was difficult to saturate the full central line
with a single pulse. For the $x=0.5$ sample, we have used a comb
of saturating pulses with total duration much greater than $T_{1}$
in the pulse sequence $(n\times\nicefrac{\pi}{2})\cdots t\cdots\nicefrac{\pi}{2}\cdots\pi$
to saturate the central line. On the other hand, for the $x=0.7$
sample the duration of saturating pulses used is much less than $T_{1}$
since the spectral width is relatively less. Consequently, the time
dependence of the recovery of longitudinal magnetization $m(t)$ is
fitted with Eq. \ref{eq:recoveryEuKFeAS} (for the $x=0.5$ sample)
and with Eq. \ref{eq:recoveryEuKFeAS-2} (for the $x=0.7$ sample)
to extract $T_{1}$ at individual temperatures. 
\begin{equation}
1-m(t)/m_{0}=A\,(0.4\, exp(-t/T_{1})+0.6\, exp(-6t/T_{1}))\label{eq:recoveryEuKFeAS}
\end{equation}

\begin{equation}
1-m(t)/m_{0}=A\,(0.1\, exp(-t/T_{1})+0.9\, exp(-6t/T_{1}))\label{eq:recoveryEuKFeAS-2}
\end{equation}

The former equation is valid for $I=3/2$ nuclei when only the central
line is saturated with a saturation sequence with duration much greater
than $T_{1}$ \cite{Andrew-ProPhySoc-78-1961}. The latter is valid
when the saturating comb has a duration much less than $T_{1}$ \cite{Simmons-PhysRev-127-1962,Suter-JPCM-10-1998}.
The coefficient $A$ takes into account the deviation from complete
saturation.

A representative longitudinal nuclear magnetization recovery curve
for $x=0.5$ sample is shown in the inset of Fig. \ref{fig:T1inv}
along with its fit with Eq. \ref{eq:recoveryEuKFeAS}. The relaxation
rates ($1/T_{1}$) for both the samples are plotted as a function
of temperature in Fig. \ref{fig:T1inv}. We have also shown the relaxation
rate ($1/T_{1}$) of Ba$_{0.3}$K$_{0.7}$Fe$_{2}$As$_{2}$ measured
with $H\parallel ab$ taken from Ref. \cite{Zhang-PRB-81-2010-Ba0.3K0.7Fe2As2-NMR}
and of Ba$_{0.6}$K$_{0.4}$Fe$_{2}$As$_{2}$ and Ba$_{0.45}$K$_{0.55}$Fe$_{2}$As$_{2}$
taken from Ref. \cite{Fukazawa-PhysicaC-470-2010}. These materials
can be considered as Ba (non magnetic) analogs of our Eu-based samples.
The relaxation rates of our Eu-based samples are more than two times
larger than those of the Ba based samples. It seems evident that the
higher relaxation rate arises from a coupling to the fluctuating Eu$^{2+}$
moments in our samples. Due to the same reason the rate is higher
for the $x=0.5$ sample which contains more Eu than the $x=0.7$ sample.
There is no signature of SDW transition or Eu$^{2+}$ moments ordering
in our $T_{1}$ measurements. For both the samples $1/T_{1}$ is almost
independent of temperature at high temperature but drops at low temperature
indicating the superconducting transition taking place in these samples.
The superconducting gap slows down the relaxation process while the
Eu$^{2+}$ moments fluctuations tries to increase the rate. Probably
the interplay of these two makes a broad transition. Similar behavior
is also reported in case of Eu$_{0.2}$Sr$_{0.8}$(Fe$_{0.86}$Co$_{0.14}$)$_{2}$As$_{2}$
\cite{Sarkar-JPCM-24-2012}. 

\begin{figure}
\centering{}\includegraphics[scale=0.32]{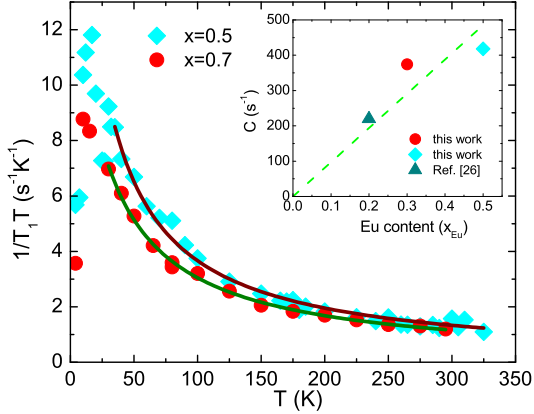}\caption{\label{fig:T1TinvCW} $1/T_{1}T$ is shown as a function of temperature.
The solid lines are Curie\textendash{}Weiss fits (described in the
text). Inset: Curie constant $C$ plotted as a function of Eu content
$x_{\mathrm{Eu}}$. The data point for $x_{\mathrm{Eu}}=0.2$ is for
Eu$_{0.2}$Sr$_{0.8}$(Fe$_{0.86}$Co$_{0.14}$)$_{2}$As$_{2}$ taken
from Ref. \cite{Sarkar-JPCM-24-2012}. The dashed line is a guide
to eye. }
\end{figure}

In Fig. \ref{fig:T1TinvCW}, we have plotted $1/T_{1}T$ as a function
of temperature and fitted with the CW law ($\frac{1}{T_{1}T}=\frac{C}{(T-\theta)}$)
in the temperature range $\geqslant30$\,K (shown as solid lines).
The fitting yields $C=374$\,s$^{-1}$ and $\theta=-22$\,K for
$x=0.7$ sample and $C=418$\,s$^{-1}$ and $\theta=-14$\,K for
the $x=0.5$ sample. It has been claimed that the Curie-Weiss behavior
of $1/T_{1}T$ indicates that the spin-lattice relaxation process
is dominated by $2$D antiferromagnetic spin fluctuations \cite{Sarkar-JPCM-24-2012}.
However it is apparent that the relaxation here is driven by fluctuations
of the Eu$^{2+}$ moments (which are three dimensional) and have little
to do with the dynamical susceptibility of the FeAs planes. In fact,
we see from the inset of Fig. \ref{fig:T1TinvCW} that the Curie constant
in $1/T_{1}T$ decreases linearly with decreasing Eu content and extrapolates
to zero for $x=0$. In reality, the $1/T_{1}$ of Ba$_{0.3}$K$_{0.7}$Fe$_{2}$As$_{2}$
(Eu-free analog) appears to vary almost linearly with $T$ \cite{Zhang-PRB-81-2010-Ba0.3K0.7Fe2As2-NMR}
which means that $1/T_{1}T$ will be nearly $T$-independent. This
further supports the fact that the relaxation process in our samples
is dominated by the Eu$^{2+}$ moments fluctuations. Indeed, the spin-lattice
relaxation rate is given by \cite{Moriya-JPSJ-1963}

\begin{equation}
1/T_{1}=\frac{2k_{B}T}{N_{A}\hbar^{2}}(\frac{\gamma_{n}}{\gamma_{e}})^{2}A_{hf}^{2}\Sigma\frac{\text{\textgreek{q}}\text{''}(q,\text{\textgreek{w}})}{\omega}
\end{equation}
where $A_{hf}$ is the hyperfine coupling constant, $\chi''(q,\omega)$
is the imaginary part of the dynamical susceptibility per mole of
electronic spin at the Larmor frequency $\omega$, $k_{B}$ is the
Boltzmann constant, $\gamma_{n}$ and $\gamma_{e}$ are nuclear and
electronic gyromagnetic ratios, respectively. The summation reduces
to $\frac{\text{\textgreek{q}}(T)\text{\textgreek{t}}}{2\text{\textgreek{p}}}$
where $\chi(T)$ is the static susceptibility of the impurity moments.
If the relaxation of the impurity Eu$^{2+}$ spins is dominated by
the interaction among them, then $1/\tau=\omega/2\pi$ (with $\omega^{2}=8J_{int}^{2}zS(S+1)/3\hbar^{2}$,
where $z$ is the number of nearest neighbors with interaction $J_{int}$)
is temperature independent. This quite naturally leads to $1/T_{1}T\propto\chi(T)$
as is observed. Quantitatively speaking, we calculated $J_{int}$
considering $\theta=2zS(S+1)/3k_{B}$ (taken to be 20K) where the
number of near neighbours for each Eu ($S=1/2$) is $z=6$. This yields
$\omega\approx4.8\times10^{11}$ rad/s. The dipolar field $A_{dip}$
was calculated \cite{Moriya-JPSJ-1963} at the As site from Eu spins
by computing $7[\sqrt{2\pi}g^{2}\mu_{B}^{2}\Sigma{1/r_{i}^{6}}]^{1/2}$.
Here $r_{i}$ is the distance of the $i^{th}$ Eu atom from a given
As nucleus and the sum is over the Eu neighbours of an As nucleus.
The summation converged beyond 100 unit cells and yielded $A_{dip}=3.65$
kOe. Taking the susceptibility of EuFe$_{2}$As$_{2}$ as \cite{Ren-PRB-78-2008}
$\chi=7.58/(T+20)$ cm$^{3}$/mole, the room temperature value of
$1/T_{1}T$ was obtained to be about $18s^{-1}K^{-1}$. The observed
value of about 1 at room temperature is then reasonable considering
the smaller concentration of Eu in $x=0.5$ or 0.7. The contribution
due to the hyperfine coupling to the Eu spins is expected to be an
order of magnitude smaller than the observed value.

\section{conclusions}

We have performed $^{75}$As NMR measurements on (Eu$_{1-x}$K$_{x}$)Fe$_{2}$As$_{2}$
($x=0.38,0.5,0.7$) samples. Our bulk measurements show no superconductivity
but a SDW transition at $197$\,K for the $x=0$ sample. This SDW
transition is suppressed and superconductivity appears for the $x=0.38,0.5$
and $0.7$ samples. The properties of the samples are largely dominated
by the Eu$^{2+}$ moments. The $^{75}$As nuclei close to the Eu$^{2+}$
moments have a very short $T_{2}$ and are driven out of our measurement
window. This results in a sharp drop of signal intensity with increasing
Eu-content in the sample and no signal is found for the full Eu sample
($x=0$). The temperature variation of the NMR shift and the spectra
can be understood by considering a distribution of magnetic environments
for the $^{75}$As nuclei; some near Eu$^{2+}$ local moments and
some far away from them. The $^{75}$As spin lattice relaxation rate
($1/T_{1}$) is seen to be dominated by the dipolar coupling to the
Eu$^{2+}$ moment fluctuations and the effect of the FeAs planes is
seen only at lower temperatures upon the onset of the superconducting
transition.

\section{acknowledgment}

We thank the Department of Science and Technology, Government of India
for financial support.

\end{document}